\documentclass[twocolumn,twoside]{revtex4}
\usepackage{graphicx}
\usepackage{fancyhdr}
\usepackage{amsmath}
\usepackage{ifthen} % for conditional statements
\usepackage[hidelinks]{hyperref}
\newboolean{uprightparticles}
\setboolean{uprightparticles}{false} %True for upright particle symbols
\usepackage{xspace} 
\usepackage{upgreek}

\def\lhcb   {\mbox{LHCb}\xspace}

\def\babar  {\mbox{BaBar}\xspace}
\def\belle  {\mbox{Belle}\xspace}

\def\cleo   {\mbox{CLEO}\xspace}

\def\lhc    {\mbox{LHC}\xspace}

\def\MagUp {\mbox{\em Mag\kern -0.05em Up}\xspace}

\ifthenelse{\boolean{uprightparticles}}%
{
 
 \def\Pgamma      {\ensuremath{\upgamma}\xspace}

 \def\Peta        {\ensuremath{\upeta}\xspace}

 \def\Ppi         {\ensuremath{\uppi}\xspace}

 \def\PDelta      {\ensuremath{\Delta}\xspace}                 
 \def\PXi         {\ensuremath{\Xi}\xspace}                 
 \def\PLambda     {\ensuremath{\Lambda}\xspace}                 
 \def\PSigma      {\ensuremath{\Sigma}\xspace}                 
 \def\POmega      {\ensuremath{\Omega}\xspace}                 
 \def\PUpsilon    {\ensuremath{\Upsilon}\xspace}

 \def\PB      {\ensuremath{\mathrm{B}}\xspace}                 
                  
 \def\PD      {\ensuremath{\mathrm{D}}\xspace}

 \def\PK      {\ensuremath{\mathrm{K}}\xspace}

 \def\Pi      {\ensuremath{\mathrm{i}}\xspace}

 \def\Ps      {\ensuremath{\mathrm{s}}\xspace}

 \def\thebaroffset{0.0em}
}
{
 
 \def\Pgamma      {\ensuremath{\gamma}\xspace}

 \def\Peta        {\ensuremath{\eta}\xspace}

 \def\Ppi         {\ensuremath{\pi}\xspace}

 \mathchardef\PDelta="7101
 \mathchardef\PXi="7104
 \mathchardef\PLambda="7103
 \mathchardef\PSigma="7106
 \mathchardef\POmega="710A
 \mathchardef\PUpsilon="7107
                  
 \def\PB      {\ensuremath{B}\xspace}                 
                  
 \def\PD      {\ensuremath{D}\xspace}

 \def\PK      {\ensuremath{K}\xspace}

 \def\Pi      {\ensuremath{i}\xspace}

 \def\Ps      {\ensuremath{s}\xspace}

 \def\thebaroffset{0.18em}
}
\newcommand{\offsetoverline}[2][\thebaroffset]{\kern #1\overline{\kern -#1 #2}}%

\makeatletter
  \newcommand{\miniscule}{\@setfontsize\miniscule{4}{5}}% \tiny: 5/6
\makeatother

\DeclareRobustCommand{\optbar}[1]{\shortstack{{\miniscule (\rule[.5ex]{1.25em}{.18mm})}
  \\ [-.7ex] $#1$}}

\def\g      {{\ensuremath{\Pgamma}}\xspace}

\def\squark    {{\ensuremath{\Ps}}\xspace}

\def\pion   {{\ensuremath{\Ppi}}\xspace}
\def\piz    {{\ensuremath{\pion^0}}\xspace}
\def\pip    {{\ensuremath{\pion^+}}\xspace}
\def\pim    {{\ensuremath{\pion^-}}\xspace}
\def\pipm   {{\ensuremath{\pion^\pm}}\xspace}

\def\kaon    {{\ensuremath{\PK}}\xspace}
\def\KorKbar {\kern \thebaroffset\optbar{\kern -\thebaroffset \PK}{}\xspace}

\def\Kp      {{\ensuremath{\kaon^+}}\xspace}
\def\Km      {{\ensuremath{\kaon^-}}\xspace}

\newcommand{\etaz}{\ensuremath{\Peta}\xspace}
\newcommand{\etapr}{\ensuremath{\Peta^{\prime}}\xspace}

\def\Dbar    {{\ensuremath{\offsetoverline{\PD}}}\xspace}
\def\D       {{\ensuremath{\PD}}\xspace}

\def\DorDbar {\kern \thebaroffset\optbar{\kern -\thebaroffset \PD}\xspace}
\def\Dz      {{\ensuremath{\D^0}}\xspace}
\def\Dzb     {{\ensuremath{\Dbar{}^0}}\xspace}
\def\Dp      {{\ensuremath{\D^+}}\xspace}
\def\Dm      {{\ensuremath{\D^-}}\xspace}
\def\Dpm     {{\ensuremath{\D^\pm}}\xspace}

\def\DpDm    {\ensuremath{\Dp {\kern -0.16em \Dm}}\xspace}

\def\Dstarp  {{\ensuremath{\D^{*+}}}\xspace}

\def\Dsp     {{\ensuremath{\D^+_\squark}}\xspace}

\def\Dspm    {{\ensuremath{\D^{\pm}_\squark}}\xspace}

\def\B       {{\ensuremath{\PB}}\xspace}

\def\BorBbar {\kern \thebaroffset\optbar{\kern -\thebaroffset \PB}\xspace}
\def\Bz      {{\ensuremath{\B^0}}\xspace}

\def\Bd      {{\ensuremath{\B^0}}\xspace}

\def\BdorBdbar {\kern \thebaroffset\optbar{\kern -\thebaroffset \Bd}\xspace}
\def\Bu      {{\ensuremath{\B^+}}\xspace}

\def\Bp      {{\ensuremath{\Bu}}\xspace}

\def\Bs      {{\ensuremath{\B^0_\squark}}\xspace}

\def\BsorBsbar {\kern \thebaroffset\optbar{\kern -\thebaroffset \Bs}\xspace}

\def\Y#1S{\ensuremath{\PUpsilon{(#1S)}}\xspace}

\def\LorLbar     {\kern \thebaroffset\optbar{\kern -\thebaroffset \PLambda}\xspace}

\newcommand{\decay}[2]{\ensuremath{#1\!\to #2}\xspace} 

\def\to                 {\ensuremath{\rightarrow}\xspace}

\def\CP                {{\ensuremath{C\!P}}\xspace}

\def\AT#1     {\ensuremath{A_{\mathrm{T}}^{#1}}\xspace}           % 2

\def\C#1      {\ensuremath{\mathcal{C}_{#1}}\xspace}                       % 9
\def\Cp#1     {\ensuremath{\mathcal{C}_{#1}^{'}}\xspace}                    % 7
\def\Ceff#1   {\ensuremath{\mathcal{C}_{#1}^{\mathrm{(eff)}}}\xspace}        % 9  
\def\Cpeff#1  {\ensuremath{\mathcal{C}_{#1}^{'\mathrm{(eff)}}}\xspace}       % 7
\def\Ope#1    {\ensuremath{\mathcal{O}_{#1}}\xspace}                       % 2
\def\Opep#1   {\ensuremath{\mathcal{O}_{#1}^{'}}\xspace}                    % 7

\newcommand{\ket}[1]{\ensuremath{|#1\rangle}}              % {b}
\newcommand{\aunit}[1]{\ensuremath{\text{\,#1}}}       
\newcommand{\tev}{\aunit{Te\kern -0.1em V}\xspace}
\newcommand{\gev}{\aunit{Ge\kern -0.1em V}\xspace}
\newcommand{\mev}{\aunit{Me\kern -0.1em V}\xspace}
\newcommand{\kev}{\aunit{ke\kern -0.1em V}\xspace}
\newcommand{\ev}{\aunit{e\kern -0.1em V}\xspace}
 
\newcommand{\mevc}{\ensuremath{\aunit{Me\kern -0.1em V\!/}c}\xspace}
\newcommand{\gevc}{\ensuremath{\aunit{Ge\kern -0.1em V\!/}c}\xspace}
\newcommand{\mevcc}{\ensuremath{\aunit{Me\kern -0.1em V\!/}c^2}\xspace}
\newcommand{\gevcc}{\ensuremath{\aunit{Ge\kern -0.1em V\!/}c^2}\xspace}
\def\fb   {\ensuremath{\aunit{fb}}\xspace}
\def\invfb   {\ensuremath{\fb^{-1}}\xspace}

\newcommand{\chisq}{\ensuremath{\chi^2}\xspace}

\def\gsim{{~\raise.15em\hbox{$>$}\kern-.85em
          \lower.35em\hbox{$\sim$}~}\xspace}
\def\lsim{{~\raise.15em\hbox{$<$}\kern-.85em
          \lower.35em\hbox{$\sim$}~}\xspace}

\def\tell1  {TELL1\xspace}
\def\ukl1   {UKL1\xspace}

\def\CPV{{\ensuremath{C\!P\!V}}\xspace}

\begin{document}
	
	\title{Mixing and \CP violation in Charm decays at LHCb}
	
	
	\author{Roberto Ribatti$^\dagger$\\
		on behalf of the LHCb collaboration}
	\affiliation{$^\dagger$Scuola Normale Superiore and INFN Sezione di Pisa, Pisa, Italy}

	\begin{abstract}
		Two measurements of mixing and \CP violation in charm decays performed at the \lhcb experiment are presented.
		The former is a measurement of the mixing observable $y_\CP-y_\CP^{\kaon\pion}$ in two-body \Dz decays, while the latter is a search for direct \CP violation in $\D^+_{(s)}\to\eta^{(\prime)}\pip$ decays.
	\end{abstract}
	
	\maketitle
	
	\thispagestyle{fancy}
	
	\section{Introduction}
	In the Standard Model (SM), mixing and \CP violation in charm are highly suppressed, due to the smallness of the CKM matrix elements involved in these processes and to the GIM mechanism.
	Measurements of mixing and \CPV in charm provide good sensitivity to beyond SM operators that couples to up-type quarks only.
	
	The \Dz mass eigenstates can be written as linear combinations of flavour eigenstates 
	$\ket{D_{1,2}} = p \ket{\Dz} \pm q\ket{\Dzb}$,
	where $p$ and $q$ are complex parameters.
	Oscillations are characterized by two parameters: the differences in mass, $x \equiv (m_1 - m_2)/\Gamma$, and in decay width, $y \equiv (\Gamma_1-\Gamma_2)/2\Gamma$, between the $D^0$ mass eigenstates (here $\Gamma$ is the average decay width of the \Dz mass eigenstates). 
	If \CP is violated, then the oscillation probabilities for \Dz and \Dzb mesons can differ.
	
	In recent years significant progress has been made in this research field, mainly due to the large samples of charmed hadron decays collected by the \lhcb experiment.
	The first observation of a non-zero value of the $y$ parameter dates back to 2010 by \babar and \belle collaborations~\cite{HFLAV:2010pgm} and today we are approaching a 3\% relative precision on its world-averaged value~\cite{LHCb:2021dcr}.
	The first observation of a non-zero value of $x$, instead, was achieved only in 2021 from the \lhcb collaboration~\cite{LHCb:2021ykz}.
	Moreover, in 2019 the \lhcb collaboration announced the first observation of \CP violation in charm decays~\cite{LHCb:2019hro}, achieved looking at the difference of the time-integrated \CP asymmetries of \decay{\Dz}{\Kp\Km} and \decay{\Dz}{\pip\pim} decays, $\Delta A_{\CP} \equiv A_{\CP}(\Kp\Km) - A_{\CP}(\pip\pim)$.
	However, the large theoretical uncertainty on nonperturbative QCD effects, which might contribute to enhance \CP violation, does not allow to unequivocally establish if this observation is compatible or not with the SM.
	Further measurements are crucial to shed light on \CP violation theoretical interpretation in the up-type quarks sector.
	
	In these proceedings, two recent \lhcb measurements of mixing and \CP violation in charm decays are presented: a measurement of the mixing observable $y_\CP-y_\CP^{\kaon\pion}$ in two-body \Dz meson decays and a search for direct \CP violation in $\D^+_{(s)}\to\eta^{(\prime)}\pip$ decays.
	
	\section{Measurement of the charm mixing parameter $y_\CP$ with $\Dz\to h^+h^-$ decays}
\begin{figure}[b]
		\centering
		\includegraphics[width=75mm]{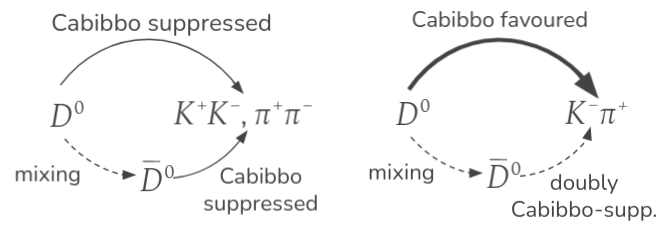}
		\caption{Sketches of the interfering amplitudes of \Dz two-body decays.} \label{fig:sketch_decay}
	\end{figure}

	The $y_\CP$ mixing observable is defined based on $\Dz$ two-body decay rates.
	As shown in Fig.~\ref{fig:sketch_decay}, the self-conjugate final states $\Dz\to f$, with $f = \Km\Kp$ or $\pim\pip$, undergo a Cabibbo-suppressed decay either directly of after mixing, while the $\Km\pip$ final state is dominated by the Cabibbo-favoured amplitude, since the contribution following mixing is suppressed by the subsequent doubly Cabibbo-suppressed decay. 
	The time-dependent decay rates can be described by exponential functions with an effective decay width. The non-zero value of $y$ introduces a shift between the two decay modes. The departure from unity of the ratio of the effective decay widths of $\Dz \to f$ and $\Dz\to\Km\pip$ is the target observable:
	\begin{align}
		&\frac{\Gamma(\Dz \to f) + \Gamma (\Dzb \to f)}{\Gamma(\Dz \to \Km\pip) + \Gamma(\Dzb \to \Kp\pim)} - 1 \simeq \\
		\nonumber &\hspace{4cm}\simeq y_{\CP}^{f} - y_{\CP}^{\kaon\pion}.
	\end{align}
	At the current experimental sensitivity, final-state dependent contributions and \CP violation effects can be neglected, hence $y_{\CP}^{f}$ is equal to $y$. 
	The correction term $y_\CP^{\kaon\pion}$ is approximately equal to $\sqrt{R_D} y \approx 6\% \,y$,~\cite{Pajero:2021jev}
	where $R_D$ is the ratio of the branching fractions of the doubly Cabibbo-suppressed $\Dz \to \Kp\pim$ decay to the Cabibbo-favoured $\Dz \to \Km\pip$ decay.
	
	The observable of interest is measured through an exponential fit to the ratio
	\begin{align}
		\label{eq:experimental_observable}
		&R^f(t) = \frac{N(\Dz\to f, t)}{N(\Dz \to \Km\pip,t)} \propto\\ \nonumber&\propto e^{-(y_{\CP}^f-y_{\CP}^{\kaon\pion}) t/\tau_{\Dz}} \frac{\epsilon(f,t)}{\epsilon(\Km\pip,t)},
	\end{align}
	where $\tau_{\Dz}$ is the lifetime of the $\Dz$ meson and $\epsilon(X)$ is the time-dependent efficiency of the $X$ decay mode.
	
	This measurement is performed using proton-proton ($pp$) collision collected at the LHCb experiment at a centre-of-mass energy of 13\tev during Run~2 (2015--2018), corresponding to an integrated luminosity of 6\invfb.
	The \Dz meson is required to come from a $\Dstarp \to \Dz \pip$ decay, such that the flavour of the $\Dz$ is determined from the charge of the slow pion.
	
	\paragraph*{Efficiency equalization}
	Trigger requirements are applied to the \Dz daughters tracks, on kinematic observables such as momentum and pseudorapidity. Applying these requirement to different final states, with different kinematic distributions, introduces discrepancies between the time-dependent efficiency of these final states.
	In order to equalize these efficiencies, a procedure called \textit{kinematic matching} is applied.
	It consists in computing new kinematic variables, called \textit{matched} variables, designed to have the same distribution for the \Dz\to$f$ decays and \Dz\to\Km\pip decay, so that if a selection on these variables is applied, the efficiency is equal for both final states.
	To compute the matched kinematic variables for a \Dz\to\Km\Kp decay, the momentum of the daughter particles in the \Dz centre-of-mass frame, which is 791\mevc for a \Dz\to\Km\Kp decay, is changed to 861\mevc (the value for a \Dz\to\Km\pip decay).
	
\begin{figure}[t]
		\centering
		\includegraphics[width=70mm]{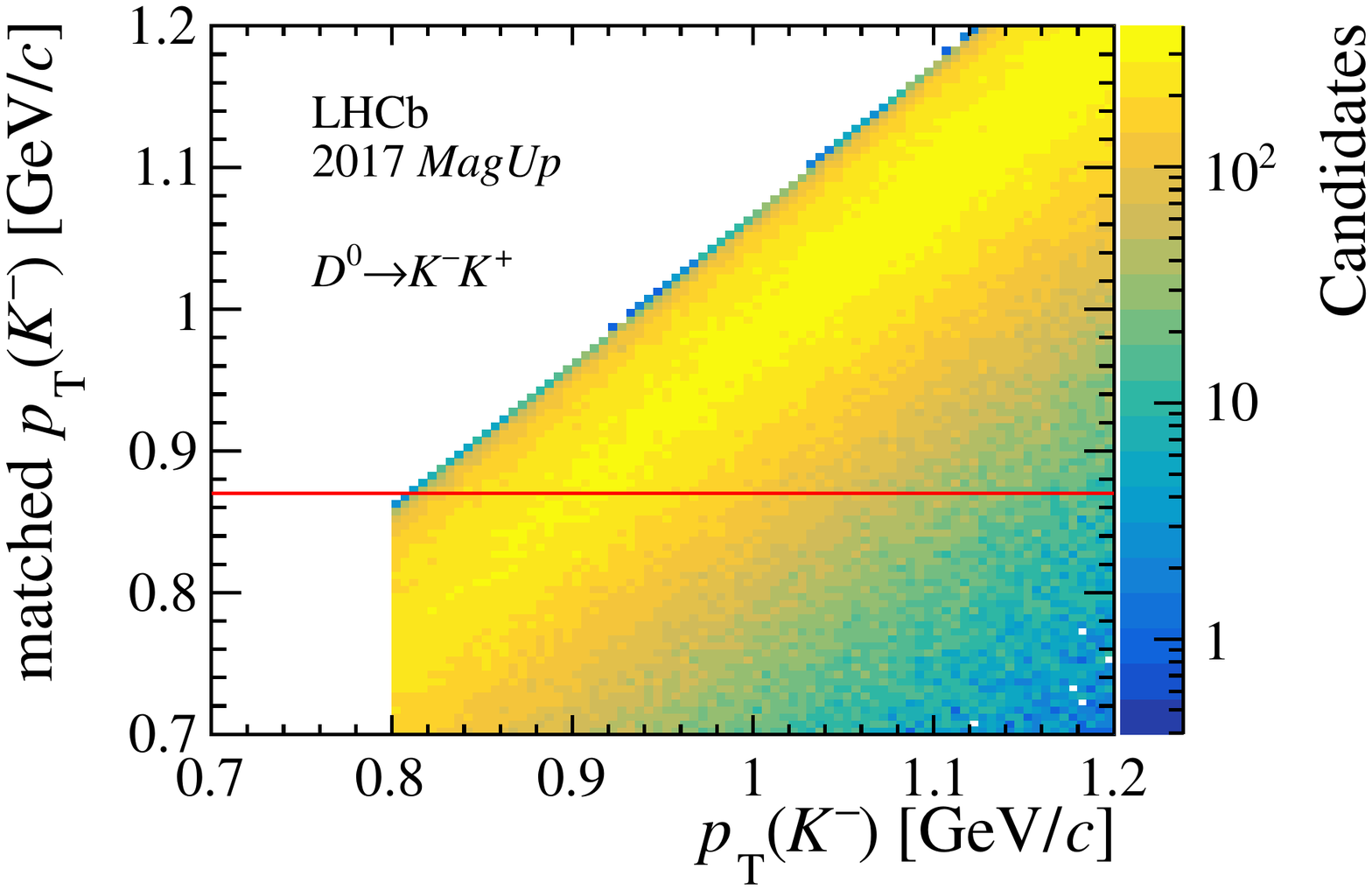}
		\caption{Matched versus original transverse momenta of a kaon from a \Dz\to\Km\Kp decay. Candidates below the red line are rejected.
			Figure from ref.~\cite{LHCb:2022gnc}.} \label{fig:ptVSmatchedpt}
	\end{figure}

	To remove the effect of the trigger requirements, tighter requirements are applied on the matched kinematic variables. Figure~\ref{fig:ptVSmatchedpt} shows an example of how the new thresholds are determined to ensure that both \Dz\to\Km\Kp and \Dz\to\Km\pip decays are selected with the same efficiency.
	
	The different interaction with the matter of kaons compared to pions, as well as particle identification requirements, produce efficiency discrepancies that cannot be removed by the kinematic matching.
	To remove these residual discrepancies, the $p$, $p_T$ and $\eta$ distributions of the $\Dstarp$ mesons of $\Dz \to f$ decays are weighted to the corresponding distributions of $\Dz \to \Km\pip$ decay, using a gradient-boosted reweighting algorithm from the \texttt{hep\_ml} library~\cite{Rogozhnikov:2016bdp}.
	
	This efficiency equalization is validated with simulated samples and also on data, through a cross-check observable,
	\begin{align}
		& R^{CC}(t) = \frac{N(\Dz\to \pim\pip, t)}{N(\Dz \to \Km\Kp,t)} \propto\\
		\nonumber&\propto e^{-y^{CC}_{\CP} t/\tau_{\Dz}} \frac{\epsilon(\pim\pip,t)}{\epsilon(\Km\Kp,t)},
	\end{align}
	where $y^{CC}_{\CP}$ is expected to be zero in the SM and is measured to be $y^{CC}_{\CP} = (-0.44\pm0.53)\times 10^{-3}$.
	
	\paragraph*{Yield determination and main backgrounds}
	The combinatorial background, mainly due to random slow pion, is the main background in the data sample. In order to disentangle the signal yield from this background, a binned \chisq fit is performed, in bins of \Dz decay time, to the $\Delta m = m(h^-h^+\pi^+_{\text{tag}})-m(h^-h^+)$ observable, where $m(h^-h^+\pi^+_{\text{tag}})$ is the mass of the $\Dstarp$ candidate and $m(h^-h^+)$ is the mass of the \Dz candidate. Most of the uncertainty related to the \Dz mass resolution is cancelled in the subtraction, considerably improving the mass resolution.
	The signal model consists of the sum of three Gaussian and of a Johnson SU~\cite{Johnson:1949} functions, while the combinatorial background is modelled with this empirical function
	\begin{align}
		&\mathcal{P}_{\text{BKG}}(\Delta m |m_0, \alpha) \equiv\\
		\nonumber&\equiv \frac{1}{\mathcal{I}_B}\cdot \sqrt{\frac{\Delta m^2}{m_0^2}-1}\cdot \exp\left(-\alpha\left(\frac{\Delta m^2}{m_0^2}-1\right)\right).
	\end{align}
	The signal yields, integrated over all decay-time bins, amount to 70 million, 18 million, and 6 million decays, respectively for the \Dz\to\Km\pip, \Dz\to\Km\Kp and \Dz \to \pim\pip decay channels.
	
	The second main background contribution comes from \Dstarp mesons that are not produced at the interaction vertex, but that rather come from a \Bz or \Bp decay.
	The reconstructed decay time of those candidates is biased toward larger values, diluting the mixing effects.
	To reduce this background, a requirement on the \Dz impact parameter with respect to the its interaction vertex, is applied.
	The fraction of residual background is fitted with simulated templates in the 2D-variables space of the \Dz impact parameter and decay time. Then the decay-time bias from the residual background is estimated with the \lhcb simulation, and subtracted.
	\begin{figure*}[t]
		\centering
		\includegraphics[width=65mm]{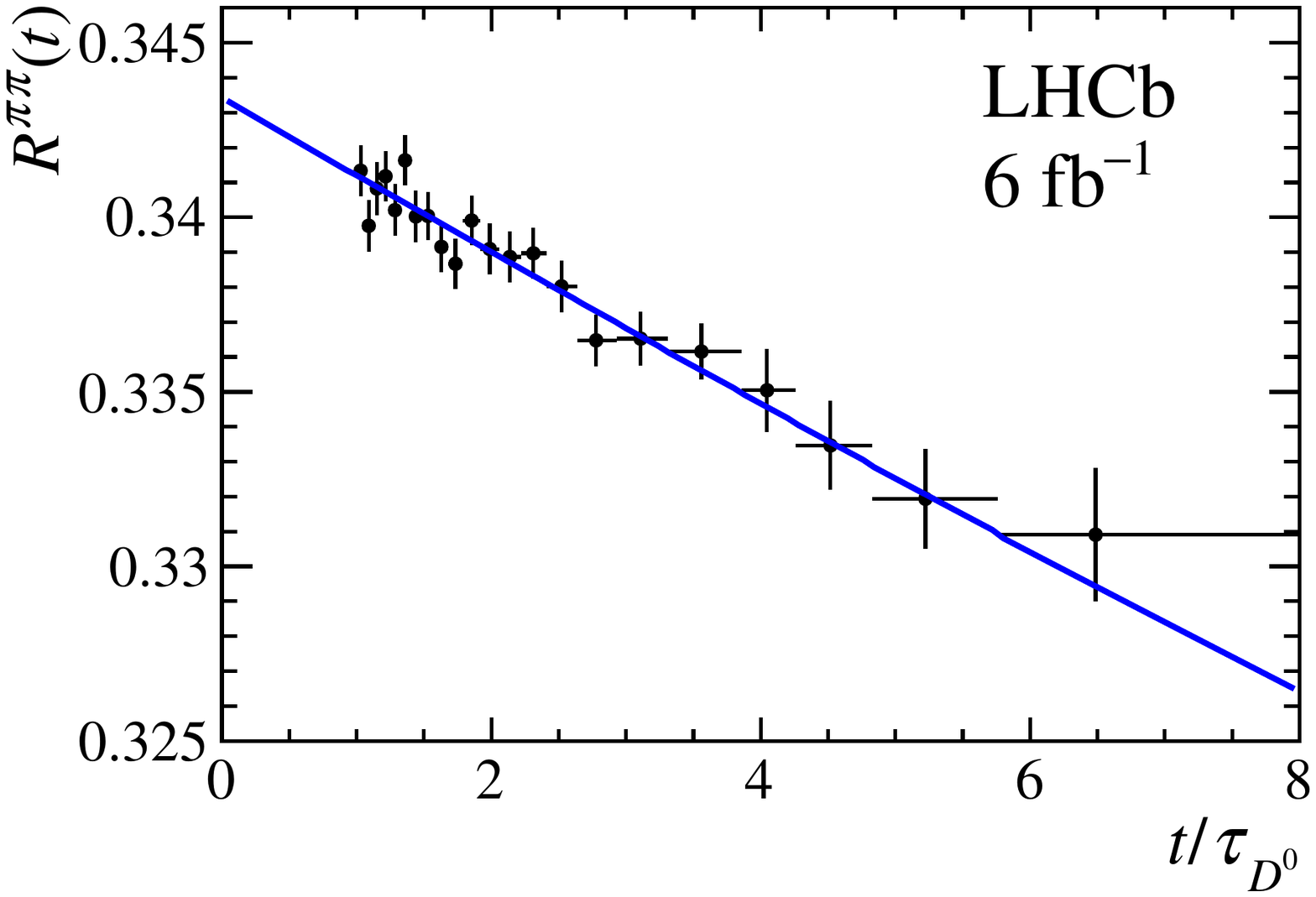}
		\includegraphics[width=65mm]{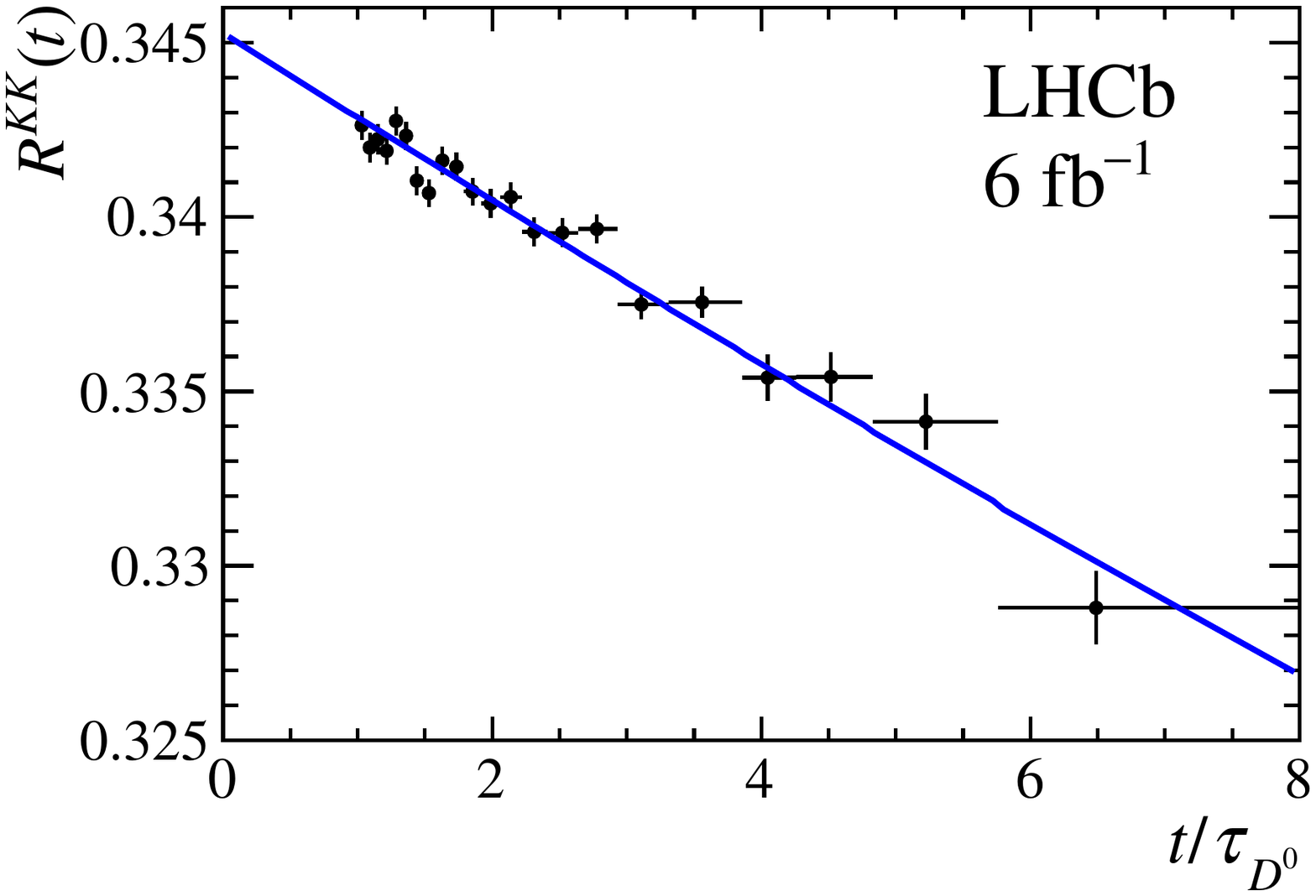}
		\caption{Corrected values of $R^{\pion\pion}(t)$ (top) and $R^{\kaon\kaon}(t)$ (bottom) for the full dataset with the fit results overlaid. Figure from ref.~\cite{LHCb:2022gnc}.} \label{fig:y_cp_results}
	\end{figure*}
	\paragraph*{Final results}
	The parameters are determined from a \chisq fit to the time-dependent $R^f(t)$ ratios. Figure~\ref{fig:y_cp_results} shows the distributions of $R^f(t)$ for the full dataset, with the fit result overlaid. The final results are
	\begin{align*}
		y_\CP^{\pion\pion}-y_\CP^{\kaon\pion} &= (6.57\pm0.53\pm0.16)\times 10^{-3}\\
		y_\CP^{\kaon\kaon}-y_\CP^{\kaon\pion} &= (7.08\pm0.30\pm0.14)\times 10^{-3},
	\end{align*}
	where the first uncertainty is statistical, while the second is systematic.
	Accounting for correlations among the systematic uncertainties, the combination is
	\begin{equation*}
		y_\CP-y_\CP^{\kaon\pion} = (6.96\pm0.26\pm0.13)\times 10^{-3}.
	\end{equation*}
	This measurement is statistically limited and it improves the world average by a factor 4.
	
	\section{Measurement of \CP asymmetries in $D_{(s)}^+\to \eta^{(\prime)}\pi^+$ decays}
	The goal of this analysis is the measurement of the time-integrated \CP asymmetry of the singly Cabibbo-suppressed \Dp\to $\eta^{(\prime)}$\pip decay, while simultaneously measuring the Cabibbo-favoured \Dsp\to $\eta^{(\prime)}$\pip decay.
	Singly Cabibbo-suppressed decay channels are of particular interest because \CP violation may manifest due to the interference of tree level amplitudes.
	No evidence for \CP violation has been observed in these decay channels by the \cleo~\cite{CLEO:2009fiz,CLEO:2013bae}, \belle~\cite{Belle:2011tmj, Belle:2021ygw} and \lhcb~\cite{LHCb:2021rou, LHCb:2017lea} collaborations. 
	
	Also this measurement is performed using $pp$-collision data collected at the LHCb experiment at a centre-of-mass energy of 13\tev during Run~2 (2015-2018), corresponding to an integrated luminosity of 6\invfb.
	Both \etaz and \etapr mesons are reconstructed in the \g\pim\pip final state. The two charged pions allow the decay vertex to be reconstructed, which is crucial to reduce the combinatorial background.
	The \CP asymmetry is defined as
	\begin{equation}
		\mathcal{A}^{\CP}(D^+_{(s)}\to f^+) \equiv \frac{\Gamma(D^+_{(s)}\to f^+)-\Gamma(D^-_{(s)}\to f^-)}{\Gamma(D^+_{(s)}\to f^+)+\Gamma(D^-_{(s)}\to f^-)},
	\end{equation}
	where $\Gamma$ is the partial decay width to the final state of interest $f$.
	For each of the analysed final states, the measured experimental observable is the raw asymmetry,
	\begin{equation}
		\mathcal{A}^{\text{raw}}(D^+_{(s)}\to f^+) \equiv \frac{N(D^+_{(s)}\to f^+)-N(D^-_{(s)}\to f^-)}{N(D^+_{(s)}\to f^+)+N(D^-_{(s)}\to f^-)},
	\end{equation}
	where $N$ refer to the measured yield.
	For small asymmetries, a first-order approximation can be used:
	\begin{align}
		\mathcal{A}^{\text{raw}}&(D^+_{(s)}\to f^+) \simeq \\
		\nonumber&\simeq\mathcal{A}^{\CP}(D^+_{(s)}\to f^+) + \mathcal{A}^{\text{prod}}(D^+_{(s)}) + \mathcal{A}^{\text{det}}(f^+).
	\end{align}
	Therefore, to extract the \CP asymmetry value from the raw asymmetry, we have to correct for the production asymmetry and the detection asymmetry. The production asymmetry is defined as
	\begin{equation}
		\mathcal{A}^{\text{prod}}(D^+_{(s)})\equiv \frac{\sigma(D^+_{(s)})-\sigma(D^-_{(s)})}{\sigma(D^+_{(s)})+\sigma(D^-_{(s)})},
	\end{equation}
	where $\sigma(D^\pm_{(s)})$ is the production cross-section of the $D^+_{(s)}$ and $D^-_{(s)}$ mesons, which differ at the \lhc, because of the \CP asymmetric initial state, $pp$.
	The detection asymmetry is defined as
	\begin{equation}
		\mathcal{A}^{\text{det}}(f^+)\equiv \frac{\epsilon(f^+)-\epsilon(f^-)}{\epsilon(f^+)+\epsilon(f^-)},
	\end{equation}
	where $\epsilon(f^\pm)$ is the detection efficiency of the final state $f^+$ and $f^-$. These two values can differ due to the different interaction with matter of particles and antiparticles.
	In our case of study, the detection asymmetry is fully due to the pion in the $D^{\pm}_{(s)}\to \eta^{(\prime)}\pipm$ decay, because the $\eta^{(\prime)}$ final state, $\g\pim\pip$ is \CP symmetric.
	In order to subtract these nuisance asymmetries, the raw asymmetry of the control channels are subtracted. These control channels have identical contributions from production and detection asymmetries, but have a negligible, or precisely known, \CP asymmetry. The control samples consist of the Cabibbo-favoured $\Dsp\to\phi\pip$ decays, where $\mathcal{A}^{\CP}$ is assumed negligible in the SM, and the singly Cabibbo-suppressed $\Dp\to\phi\pip$ decay, where $\mathcal{A}^{\CP}(\Dp\to\phi\pip) = (0.5 \pm 5.1) \times 10^{-4}$ has been previously measured by the \lhcb collaboration~\cite{LHCb:2019dwr}.
	The $\phi$ meson is reconstructed in the self-coniugate $\Km\Kp$ final state.
	
	\begin{figure}[t]
		\centering
		\includegraphics[width=90mm]{fig/Acp_Dstoetapi_pdf/Fig2.pdf}
		\includegraphics[width=90mm]{fig/Acp_Dstoetapi_pdf/Fig3.pdf}
		\caption{Distributions of $m(\eta^{(\prime)}\pip)$ (left) and $m(\eta^{(\prime)}\pip)$ (right) for candidates with $m(\g\pim\pip)$ in the range 936--976\mevcc (top) and 526--570\mevcc (bottom). The fit results are superimposed. Figure from ref.~\cite{LHCb:2022pxf}.} \label{fig:fit_ACP_metapip}
	\end{figure}
	\begin{figure}[t]
		\includegraphics[width=55mm]{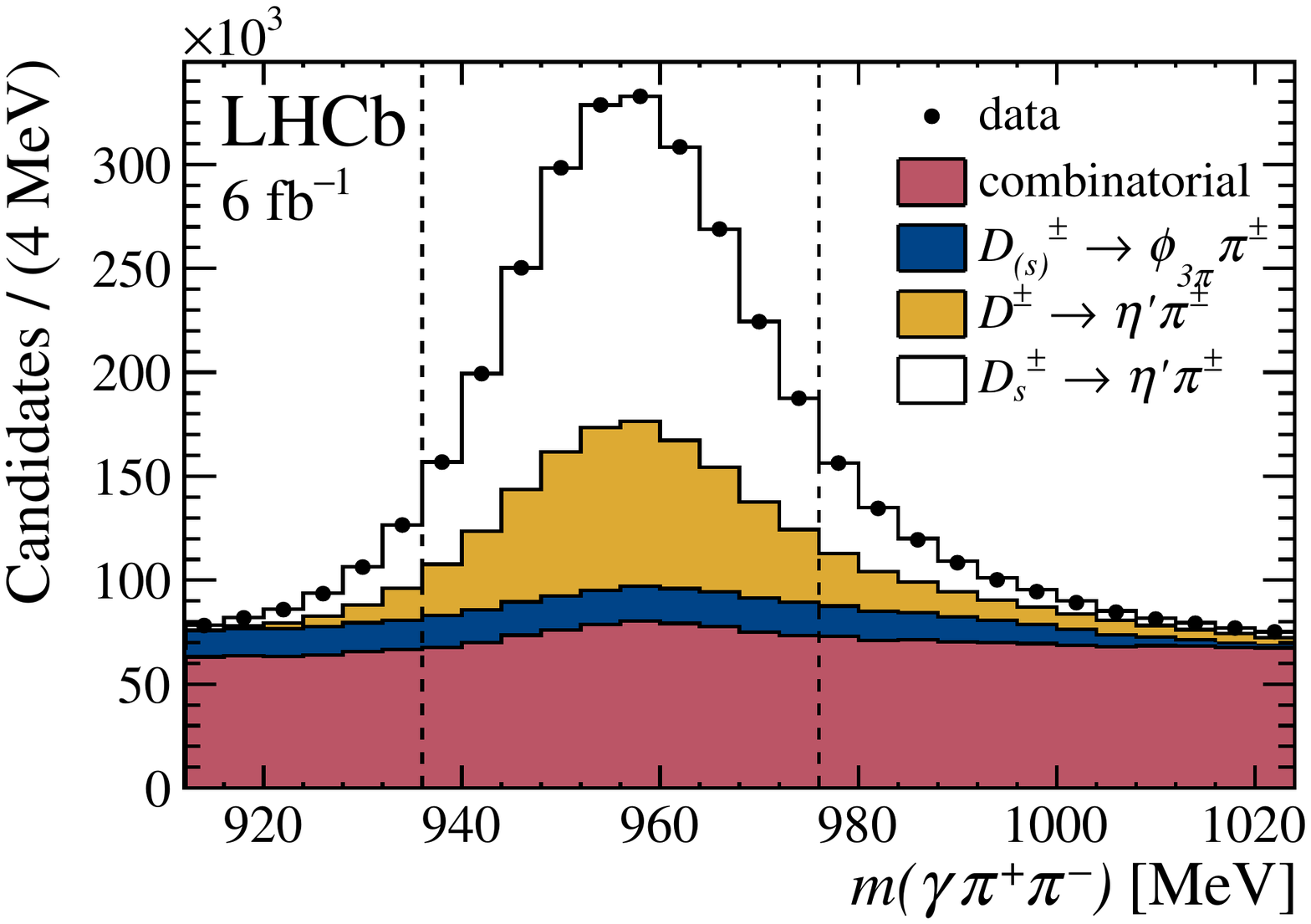}
		\includegraphics[width=55mm]{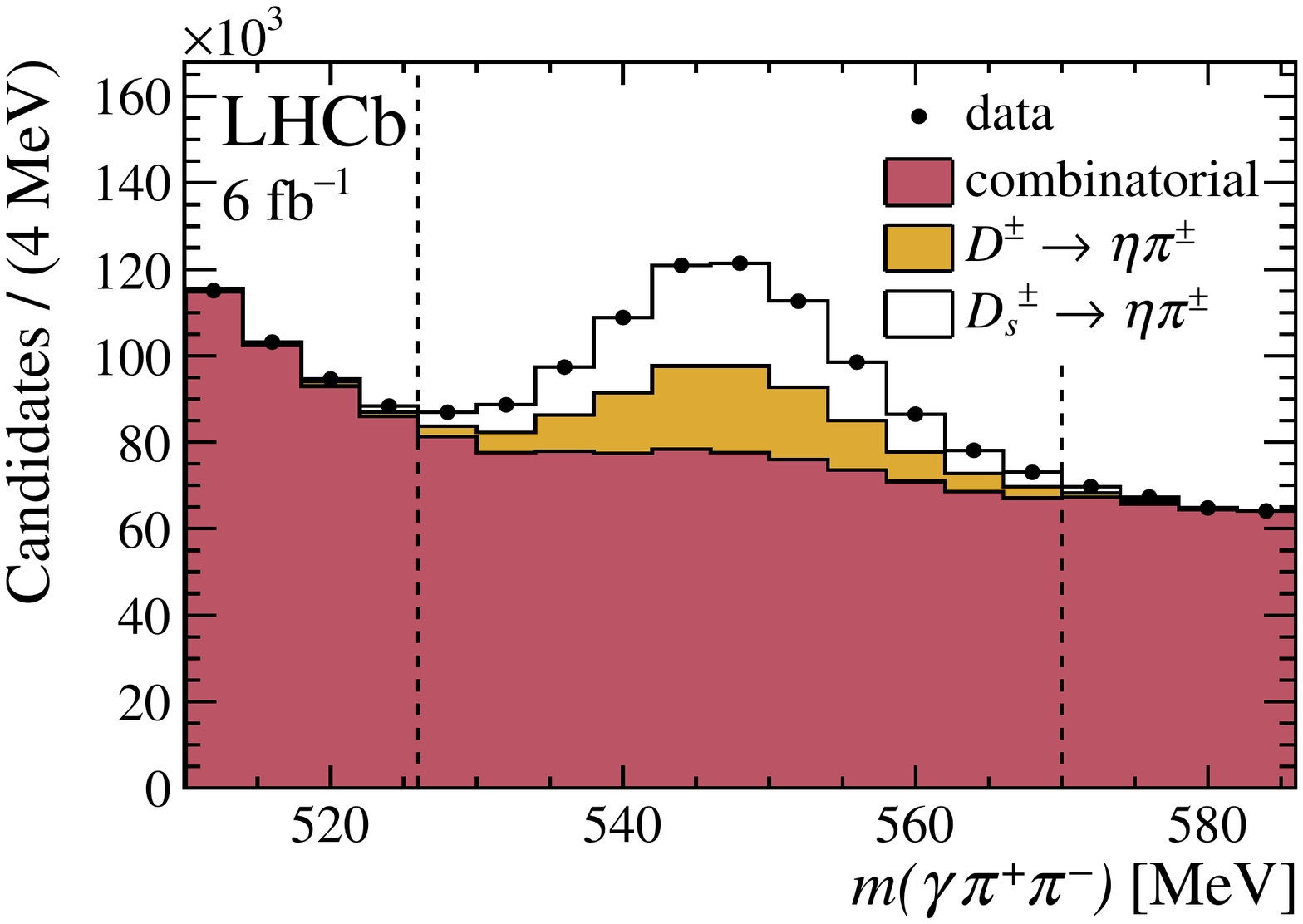}
		\caption{Distributions of $m(\g\pim\pip)$ for $D^{\pm}_{(s)}\to\etapr\pipm$ (top) and $D^{\pm}_{(s)}\to\etaz\pipm$ (bottom). The fit results are superimposed. Figure from ref.~\cite{LHCb:2022pxf}.}\label{fig:fit_ACP_mgpimpip}
	\end{figure}
	
	The experimental observables are
	\begin{align}
		\mathcal{A}^\CP&(\Dp \to \eta^{(\prime)}\pip) = \\
		\nonumber&=\mathcal{A}^{\text{raw}}(\Dp \to \eta^{(\prime)}\pip)-\mathcal{A}^{\text{raw}}(\Dp \to \phi\pip) + \\
		\nonumber&+\mathcal{A}^\CP(\Dp \to \phi\pip),\\
		\mathcal{A}^\CP&(\Dsp \to \eta^{(\prime)}\pip) = \\
		\nonumber&=\mathcal{A}^{\text{raw}}(\Dsp \to \eta^{(\prime)}\pip)-\mathcal{A}^{\text{raw}}(\Dsp \to \phi\pip).
	\end{align}
	\paragraph*{Determination of the raw asymmetry}
	The main background in the sample is combinatorial background.
	In order to determine the raw asymmetry, signal yields need to be extracted and disentangled from this background. To do this, a binned maximum-likelihood fit is performed simultaneously in the 3D variable space $[m(\eta^{(\prime)}\pip),m(\eta^{(\prime)}\pim),m(\g\pim\pip)]$.
	Both the signal and combinatorial background are modeled with empirical functions. The \Dspm and \Dpm signal models are Johnson SU\cite{Johnson:1949} functions in the $m(\eta^{(\prime)}\pipm)$ variable. A quadratic dependence on $m(\g\pim\pip)$ is allowed for the mean and the width, while $\Delta m \equiv m(\Dsp)-m(\Dp)$ linearly depends on it. The two Johnson SU parameters that mainly determine the asymmetry and kurtosis of the distribution are shared between $\Dsp$ and $\Dp$ and between positively and negatively charged candidates.
	The model for the combinatorial background is a third-order Chebyshev polynomial in the $m(\eta^{(\prime)}\pipm)$ variable, independent in each bin of $m(\g\pim\pip)$ and for positively and negatively charged candidates.
	Only for the $D^+_{(s)}\to \etapr\pip$ decay, a further background is considered, coming from $\D^+_{(s)}\to \phi(\to\pim\pip\piz)\pip$ decays where the \piz is misreconstructed as a photon.
	The shape of this background is fixed from simulation.
	Figures~\ref{fig:fit_ACP_metapip} and \ref{fig:fit_ACP_mgpimpip} show the $m(\eta^{(\prime)}\pipm)$ and $m(\g\pim\pip)$ distributions with the fit results superimposed.
	
	\paragraph*{Control channel yields extraction}
	Due to the absence of neutral particles, the control channels have much less background.
	To correctly subtract nuisance asymmetries, the kinematics distribution (transverse momentum and pseudorapidity of $D^+_{(s)}$ and tagging pion) of the control channels have to be weighted to those of the signal channels in the 
	The small background component is subtracted using the \textit{sPlot} technique~\cite{Pivk:2004ty}, using $m(\eta^{(')\pip})$ as discriminating variable.
	
	\paragraph*{Final results}
	The final value of $\mathcal{A}^{\CP}$ after the combination with previous \lhcb measurements using different $\eta$ final states~\cite{LHCb:2021rou} or independent data samples~\cite{LHCb:2017lea} are:
	\begin{align*}
		\mathcal{A}^\CP(\Dp\to\etaz\pip) &=(0.13\pm0.50\pm0.18)\%\\
		\mathcal{A}^\CP(\Dp\to\etapr\pip) &=(0.43\pm0.17\pm0.10)\%\\
		\mathcal{A}^\CP(\Dsp\to\etaz\pip) &=(0.48\pm0.42\pm0.17)\%\\
		\mathcal{A}^\CP(\Dsp\to\etapr\pip) &=(-0.04\pm0.11\pm0.09)\%,
	\end{align*}
	where the first uncertainty is statistical, and the second is systematic.
	The results are compatible with the conservation of the \CP symmetry. For the $\Dp\to\etaz\pip$, $\Dp\to\etapr\pip$ and $\Dsp\to\etapr\pip$ decay channels, this is the most precise measurement to date.
	
	\section{Future prospects}
	We have just barely started to approach the SM upper limits for \CP violation in charm decays.
	\mbox{\lhcb--Upgrade~I} started in 2022 and over 23 (50)\invfb of integrated luminosity are expected to be collected by the end of Run~3 (4) in 2025 (2032).
	Moreover, the new \lhcb data acquisition and trigger system will help to increase the reconstruction efficiency per \invfb, in particular for hadronic charm decays~\cite{Bediaga:2018lhg}. This large amount of data will allow for a significant improvement in the accuracy of mixing and \CP-violation measurements in the charm sector.
	
	%\begin{thebibliography}{9}   % Use for  1-9  references


\begin{thebibliography}{99} % Use for 10-99 references
		
		\bibitem{HFLAV:2010pgm}
		D.~Asner \textit{et al.} [HFLAV group],
		%``Averages of $b$-hadron, $c$-hadron, and $\tau$-lepton properties,''
		arXiv:1010.1589.
		
		\bibitem{HFLAV:2022pwe}
		Y.~Amhis \textit{et al.} [HFLAV group],
		%``Averages of $b$-hadron, $c$-hadron, and $\tau$-lepton properties as of 2021,''
		arXiv:2206.07501.
		
		\bibitem{LHCb:2021dcr}
		R.~Aaij \textit{et al.} [LHCb collaboration],
		%``Simultaneous determination of CKM angle $\gamma$ and charm mixing parameters,''
		JHEP \textbf{12}, 141 (2021),
		%doi:10.1007/JHEP12(2021)141
		arXiv:2110.02350.
		
		\bibitem{LHCb:2021ykz}
		R.~Aaij \textit{et al.} [LHCb collaboration],
		%``Observation of the Mass Difference Between Neutral Charm-Meson Eigenstates,''
		Phys. Rev. Lett. \textbf{127}, no.11, 111801 (2021),
		%doi:10.1103/PhysRevLett.127.111801
		arXiv:2106.03744.
		
		\bibitem{LHCb:2019hro}
		R.~Aaij \textit{et al.} [LHCb collaboration],
		%``Observation of CP Violation in Charm Decays,''
		Phys. Rev. Lett. \textbf{122}, no.21, 211803 (2019),
		%doi:10.1103/PhysRevLett.122.211803
		arXiv:1903.08726.
		
		\bibitem{Pajero:2021jev}
		T.~Pajero and M.~J.~Morello,
		%``Mixing and CP violation in D$^{0}$ \textrightarrow{} K$^{−}$\ensuremath{\pi}$^{+}$ decays,''
		JHEP \textbf{03}, 162 (2022),
		%doi:10.1007/JHEP03(2022)162
		arXiv:2106.02014.
		
		\bibitem{LHCb:2022gnc}
		R.~Aaij \textit{et al.} [LHCb collaboration],
		%``Measurement of the charm mixing parameter~$y_{CP} - y_{CP}^{K\pi}$ using two-body $D^0$ meson decays,''
		Phys. Rev. D \textbf{105}, no.9, 092013 (2022),
		arXiv:2202.09106.
		
		\bibitem{Rogozhnikov:2016bdp}
		A.~Rogozhnikov,
		%``Reweighting with Boosted Decision Trees,''
		J. Phys. Conf. Ser. \textbf{762}, no.1, 012036 (2016).
		%doi:10.1088/1742-6596/762/1/01203[physics.data-an]].
		
		\bibitem{Johnson:1949}
		N. L. Johnson, 
		Biometrika 36, 149 (1949).
		
		\bibitem{CLEO:2009fiz}
		H.~Mendez \textit{et al.} [CLEO collaboration],
		%``Measurements of D Meson Decays to Two Pseudoscalar Mesons,''
		Phys. Rev. D \textbf{81}, 052013 (2010),
		arXiv:0906.3198.
		
		\bibitem{CLEO:2013bae}
		P.~U.~E.~Onyisi \textit{et al.} [CLEO collaboration],
		%``Improved Measurement of Absolute Hadronic Branching Fractions of the $D_s^+$ Meson,''
		Phys. Rev. D \textbf{88}, no.3, 032009 (2013),
		arXiv:1306.5363.
		
		\bibitem{Belle:2011tmj}
		E.~Won \textit{et al.} [Belle collaboration],
		%``Observation of $D^+ \rightarrow K^{+} \eta^{(\prime)}$ and Search for CP Violation in $D^+ \rightarrow \pi^+ \eta^{(\prime)}$ Decays,''
		Phys. Rev. Lett. \textbf{107}, 221801 (2011).
		arXiv:1107.0553.
		
		\bibitem{Belle:2021ygw}
		Y.~Guan \textit{et al.} [Belle collaboration],
		%``Measurement of branching fractions and $CP$ asymmetries for $D_s^{+} \rightarrow K^+ (\eta, \pi^0) $ and $D_s^{+} \rightarrow \pi^+ (\eta, \pi^0)$ decays at Belle,''
		Phys. Rev. D \textbf{103}, 112005 (2021),
		arXiv:2103.09969.
		
		\bibitem{LHCb:2021rou}
		R.~Aaij \textit{et al.} [LHCb collaboration],
		%``Search for CP violation in $ {D}_{(s)}^{+}\to {h}^{+}{\pi}^0 $ and $ {D}_{(s)}^{+}\to {h}^{+}\eta $ decays,''
		JHEP \textbf{06}, 019 (2021),
		%doi:10.1007/JHEP06(2021)019
		arXiv:2103.11058.
		
		\bibitem{LHCb:2017lea}
		R.~Aaij \textit{et al.} [LHCb collaboration],
		%``Measurement of $CP$ asymmetries in $D^{\pm}\rightarrow \eta^{\prime} \pi^{\pm}$ and $D_s^{\pm}\rightarrow \eta^{\prime} \pi^{\pm}$ decays,''
		Phys. Lett. B \textbf{771}, 21-30 (2017),
		%doi:10.1016/j.physletb.2017.05.013
		arXiv:1701.01871.
		
		\bibitem{LHCb:2019dwr}
		R.~Aaij \textit{et al.} [LHCb collaboration],
		%``Search for $CP$ violation in $D_s^+\to K_S^0 \pi^+$, $D^+\to K_S^0 K^+$ and $D^+\to \phi \pi^+$ decays,''
		Phys. Rev. Lett. \textbf{122}, no.19, 191803 (2019),
		%doi:10.1103/PhysRevLett.122.191803
		arXiv:1903.01150.
		
		\bibitem{Pivk:2004ty}
		M.~Pivk and F.~R.~Le Diberder,
		%``SPlot: A Statistical tool to unfold data distributions,''
		Nucl. Instrum. Meth. A \textbf{555}, 356-369 (2005),
		%doi:10.1016/j.nima.2005.08.106
		arXiv:physics/0402083.
		
		\bibitem{LHCb:2022pxf}
		R.~Aaij \textit{et al.} [LHCb collaboration],
		%``Measurement of $CP$ asymmetries in $D^+_{(s)}\rightarrow \eta \pi^+$ and $D^+_{(s)}\rightarrow \eta^{\prime} \pi^+$ decays,''
		arXiv:2204.12228.
		
		\bibitem{Bediaga:2018lhg}
		R. Aaij \textit{et al.} [LHCb collaboration], 
		arXiv:1808.08865.
		
	\end{thebibliography}
\end{document}